\documentstyle[a4,12pt]{article}
\def\be{\begin{equation}}
\def\ee{\end{equation}}
\def\ba{\begin{eqnarray}}
\def\ea{\end{eqnarray}}
\def\mref#1{Eq.(\ref{Eq:#1})}

\def\mlab#1{\label{Eq:#1}}

\def\PR#1#2#3 {{\it Phys. Rev. }{\bf D#1} #2 {(#3)} }
\def\PRL#1#2#3 {{\it Phys. Rev. Lett. }{\bf #1} #2 {(#3)} }
\def\PL#1#2#3 {{\it Phys. Lett. }{\bf #1} #2 {(#3)}  }
\def\AP#1#2#3 {{\it Ann, Phys. }{\bf #1} #2 {(#3)} }
\def\ZP#1#2#3 {{\it Z. Phys. }{\bf #1} #2 {(#3)} }
\def\NP#1#2#3 {{\it Nucl. Phys. }{\bf #1} #2 {(#3)}  }
\def\MPL#1#2#3 {{\it Mod. Phys. Lett.}{\bf #1} #2 {(#3)}  }
\def\NC#1#2#3 {{\it Nuov. Cimm. }{\bf #1} #2 {(#3)}  }
\def\PREP#1#2#3 {{\it Phys. Report }{\bf #1} #2 {(#3)}  }
\def\PROG#1#2#3 {{\it Prog. Theor. Phys. }{\bf #1} #2 {(#3)}   }

\def\sq2{{1\over{\sqrt{2}}}}

\def\g5{\gamma_5}

\begin{document}
\parskip=5pt plus 1pt minus 1pt

\begin{flushright}
{DPNU-96-36} \\
{July 1996}
\end{flushright}

\vspace{0.2cm}

\begin{center}
{\Large\bf Getting at the Quark Mass Matrices}
\end{center}

\vspace{0.3cm}

\begin{center}
{\sc K. Harayama}\footnote{Electronic address: harayama@eken.phys.nagoya-u.ac.jp} , 
{\sc N. Okamura}\footnote{Electronic address: okamura@eken.phys.nagoya-u.ac.jp} ,
{\sc A.I. Sanda}\footnote{Electronic address: sanda@eken.phys.nagoya-u.ac.jp} ,
{\sc Zhi-zhong Xing}\footnote{Electronic address: xing@eken.phys.nagoya-u.ac.jp}
\end{center}
\begin{center}
{\it Department of Physics, Nagoya University, Chikusa-ku, Nagoya 464-01, Japan}
\end{center}

\vspace{2.5cm}

\begin{abstract}
We present a class of {\it Ans$\it\ddot{a}$tze} for the up and down quark  mass matrices which leads
approximately to: 
$|V_{us}| \sim \sqrt{m_d / m_s}$,
$|V_{cb}| \sim m_s / m_b$, and 
$|V_{ub} / V_{cb}| \sim \sqrt{m_u / m_c}$.
Sizes of the Kobayashi-Maskawa matrix elements are controlled solely by quark mass ratios.
In particular, we introduce no other small parameter and our results do not
rely on delicate cancellation.
\end{abstract}

\newpage

\section{Introduction}

Within the context of the standard electroweak model, the origin of quark masses 
and the Kobayashi-Maskawa (KM) flavor mixing matrix \cite{KM} remains a mystery . 
Therefore, a search for the origin of the KM matrix and quark
masses may show us how to get at the physics beyond the standard model.

There are 36 parameters in $3\times 3$ up and down quark mass matrices.
In contrast, there are 10 experimental constraints - 6 quark masses 
and 4 KM matrix elements.
As there are too many parameters,
it is often more constructive to start with a theory - predict the mass 
matrices and check them with experiments. 
In this paper, we take a different approach and present a class of {\it Ans$\it\ddot{a}$tze} for 
the mass matrices. 

Current experimental data suggest the following relations between
quark masses and the KM matrix elements \cite{PDG}:
\ba
|V_{us}| & \approx & \sqrt{\frac{m_d}{m_s}} \; \approx \; 0.22 \; , \nonumber \\
|V_{cb}| & \approx & \frac{m_s}{m_b} \; \approx \; 0.02 ~ \sim ~  0.06 \; , 
\mlab{clue} \\
\left | \frac{V_{ub}}{V_{cb}}\right | & \approx & \sqrt{\frac{m_u}{m_c}} \;
\approx \; 0.035 ~ \sim ~ 0.09 \; . \nonumber
\ea 
These relations may be an
accident. We take a view that the Nature is giving us a clue. We thus take 
them seriously, and give a wide class of mass matrices which can
lead to relationships of this form.

As mentioned above, 26 out of 36 parameters of quark mass matrices
have nothing to do with quark
masses or KM matrix elements. It is convenient to remove all but essential
parameters out of the mass matrices.
As first shown in Ref.\cite{Branco},  
it is always possible to find a weak basis 
where arbitrary $3\times 3$ quark mass matrices take the 
nearest-neighbor interaction (NNI) form:
\begin{equation}
M \; = \; m_3 M^{\prime} \; = \; m_3 \left( \begin{array}{ccc}
                                                0     	& a 	& 0     \\

                                                c 	& 0     & b	\\
                                                0       & d 	& e
                                            \end{array} \right) \; . 
\mlab{mass}
\end{equation}
Here we have rescaled all matrix elements of $M$ by use of the 3rd generation mass, $m_3$.
By redefining the phases of left-handed and right-handed 
quark fields, one can choose all elements of $M^{\prime}$ to be real and non-negative. 
In this case, a diagonal phase matrix 
\begin{equation}
P \; = \; \left (\matrix{
1	& 0	& 0 \cr
0	& \exp({\rm i}\theta_2)	& 0 \cr
0	& 0	& \exp({\rm i}\theta_3) \cr } \right ) \; 
%\mlab{Pmatrix}
\end{equation}
will enter the flavor mixing matrix. Diagonalizing $M M^{\rm T}$ by the 
orthogonal transformation
\begin{eqnarray}
O^{\rm T} M M^{\rm T} O & = & \left( \begin{array}{ccc}
             m_1^2 	& 0         	& 0         \\
             0          & m_2^{2} 	& 0         \\
             0          & 0         	& m_3^{2}
           \end{array} \right) \; ,
\mlab{Diag}
\end{eqnarray}
we can obtain the KM matrix $V$ for flavor mixings:
\begin{equation}
V \; = \; O_{\rm u}^{\rm T} P O_{\rm d} \; ,
\mlab{V}
\end{equation}
where the subscripts ``u'' and ``d'' stand for up and down, respectively.

In this analysis,  $V$ contains 12 free parameters: 6 of them are
determined from quark mass eigenvalues, and 4 from the experimental data on
$V$. Thus all but two of the parameters in $M_{\rm u}$ and $M_{\rm d}$ 
can be determined.

Now, let us require that $M_{\rm u}$ and $M_{\rm d}$ 
given in \mref{mass} reproduce correct quark mass eigenvalues.
For each quark sector we introduce 
four real (positive) parameters $p$, $q$, $y$ and $z$, which satisfy the relations
$$
p \; = \; \frac{m^2_1 + m^2_2}{m^2_3} \; , ~~~~~~~~
q^2 \; = \;  \frac{m_1 m_2}{m^2_3} \; ;
\eqno(6{\rm a})
$$
and
$$
a \; = \; \frac{q z}{y} \; , ~~~~~~~~
c \; = \; \frac{q}{y z} \; , ~~~~~~~~
e \; = \; y^2 \; .
\eqno(6{\rm b})
$$
As a result, the mass matrix elements $b$ and $d$ can be given by
\setcounter{equation}{6}
\begin{eqnarray}
b & = & \sqrt{\frac{p+1-y^{4} \pm R}{2} ~ - ~ \frac{q^2}{y^2 z^2}} \; , \nonumber \\
d & = & \sqrt{\frac{p+1-y^{4} \mp R}{2} ~ - ~ \frac{q^2 z^2}{y^2}} \; ,
\mlab{bd}
\end{eqnarray}
where
\begin{equation}
R \; \equiv \; \sqrt{\left (1+p-y^4\right )^2 - 4 \left (p+q^4\right )
+ 4 q^2 y^2 \left (z^2 +\frac{1}{z^2}\right )} \; .
\mlab{R}
\end{equation}
The quark mass matrix $M^{\prime}$ turns out to be the following form \cite{HO}: 
\begin{equation}
M^{\prime} \; = \; \left ( \begin{array}{ccc}
            0                 & \displaystyle\frac{ q z }{ y } & 0         \\
            \displaystyle\frac{ q }{ y z } & 0                 & b \\
            0                 & d        & y^{2}
         \end{array} \right) \; ,
\mlab{Mprime}
\end{equation}
whose structure is characterized by the dimensionless parameters $y$ and $z$.
Note that  $b$ and $d$ have two different solutions as given in \mref{bd}.
Subsequently, we refer the case that $b$ takes plus (minus) sign and $d$ takes minus 
(plus) sign to case (I) (case (II)).

\section{A class of {\it Ans$\it\ddot{a}$tze}}

Our {\it Ans$\it\ddot{a}$tze} are to take the NNI form for the up and down quark matrices, and 
require that their parameters satisfy the following conditions:
\be
y \; \sim \; O(1) \; , ~~~~~~~~
z \; \sim \; O(1) \; , ~~~~~~~~
\left (1-y^4 \right )^2 \; \gg \; 2p \left (1+y^4 \right ) \; .
\mlab{restriction}
\ee
The form given in \mref{Mprime} is completely general. The {\it Ans$\it\ddot{a}$tze} made in 
\mref{restriction} imply that when quark mass matrices are transformed
into the NNI form, the parameters $y$ and $z$ are restricted to
a certain region.

{\it It may be that this class of mass matrices have nothing to do with
the Nature. For that matter, singling out the NNI basis
on which we assume \mref{restriction} may lead us astray. This assumption
must be judged by its consequences}.

\section{Results}

In this approximation, we get
\begin{equation}
R \; \approx \; \left (1 - y^4 \right ) ~ - ~ \frac{1+y^4}{1-y^4} p \; .
\mlab{Rapprox}
\end{equation}
Then we obtain the magnitudes of $b$ and $d$ as follows:
$$
{\rm case  ~ (I):} \;\;\; b \; \approx \; \sqrt{1-y^4} \; , ~~~~~~~~
d \; \approx \; \sqrt{\frac{p}{1-y^4}} \; ; 
\eqno(12{\rm a})
$$
$$
{\rm case ~ (II):} \;\;\; b \; \approx \; \sqrt{\frac{p}{1-y^4}} \; , ~~~~~~~~
d \; \approx \; \sqrt{1-y^4} \; .
\eqno(12{\rm b}) 
$$
We insist that either case (I) or case (II) is chosen for both $M_{\rm u}$ and $M_{\rm d}$.
This leads to similar structure for both $M_{\rm u}$ and $M_{\rm d}$.
Also, by doing so, the diagonal elements of the KM matrix are neary unity.
Note that $b\gg d$ in case (I) is not favored by current data.
To see this point more clearly, we compute $V_{cb}$ in case (I) and obtain
\setcounter{equation}{12}
\be
V_{cb} \; \approx \; y^2_{\rm u} \sqrt{1-y^4_{\rm d}} \exp({\rm i}\theta_2)
~ - ~ y^2_{\rm d} \sqrt{1-y^4_{\rm u}} \exp({\rm i}\theta_3)
\mlab{case1}
\ee
at the lowest-order expansion in $p$ and $q$. 
We find that this result is quite unsatisfactory.
It is nearly independent of the quark mass 
ratios; and it has to involve large cancellation between the 
$\exp({\rm i}\theta_2)$ and
$\exp({\rm i}\theta_3)$ terms to agree with the measured value of $|V_{cb}|$.
We, therefore focus our attention on case(II).

By use of \mref{Rapprox} and Eq.(12b), we calculate the KM matrix $V$ by 
keeping only the leading terms in $p$ and $q$.
The diagonal elements of $V$ can be given as 
\begin{equation}
V_{ud} \; \approx \; 1 \; , ~~~~~~~~
V_{cs} \; \approx \; \exp({\rm i}\theta_2) \; , ~~~~~~~~
V_{tb} \; \approx \; \exp({\rm i}\theta_3) 
\mlab{Vdiagonal}
\end{equation}
in lowest-order approximations. We also find 
\small
\begin{eqnarray}
V_{us} & \approx & - y_{\rm d} z_{\rm d} \sqrt{\frac{m_d}{m_s}} +
y_{\rm u} z_{\rm u} \sqrt{\frac{m_u}{m_c}} \exp({\rm i}\theta_2) \; , \nonumber \\
V_{cd} & \approx & - y_{\rm u} z_{\rm u} \sqrt{\frac{m_u}{m_c}} + y_{\rm d} z_{\rm d} 
\sqrt{\frac{m_d}{m_s}} \exp({\rm i}\theta_2) \; ; \nonumber \\
V_{cb} & \approx & \frac{y^2_{\rm d}}{\sqrt{1-y^4_{\rm d}}} \frac{m_s}{m_b} \exp({\rm i}\theta_2) -
\frac{y^2_{\rm u}}{\sqrt{1-y^4_{\rm u}}} \frac{m_c}{m_t} \exp({\rm i}\theta_3) \; , \nonumber \\
V_{ts} & \approx & \frac{y^2_{\rm u}}{\sqrt{1-y^4_{\rm u}}} \frac{m_c}{m_t} \exp({\rm i}\theta_2) -
\frac{y^2_{\rm d}}{\sqrt{1-y^4_{\rm d}}} \frac{m_s}{m_b} \exp({\rm i}\theta_3) \; ; 
\mlab{Vresult} \\
V_{ub} & \approx & \frac{z_{\rm d} \sqrt{1-y^4_{\rm d}}}{y_{\rm d}} \sqrt{\frac{m_d}{m_s}}
\frac{m_s}{m_b} + y_{\rm u} z_{\rm u} \sqrt{\frac{m_u}{m_c}} \left [
\frac{y^2_{\rm d}}{\sqrt{1-y^4_{\rm d}}} \frac{m_s}{m_b} \exp({\rm i}\theta_2)
- \frac{1}{y^2_{\rm u} \sqrt{1-y^4_{\rm u}}} \frac{m_c}{m_t} \exp({\rm i}\theta_3) \right ] 
\; , \nonumber \\
V_{td} & \approx & \frac{z_{\rm u} \sqrt{1-y^4_{\rm u}}}{y_{\rm u}} \sqrt{\frac{m_u}{m_c}}
\frac{m_c}{m_t} + y_{\rm d} z_{\rm d} \sqrt{\frac{m_d}{m_s}} \left [
\frac{y^2_{\rm u}}{\sqrt{1-y^4_{\rm u}}} \frac{m_c}{m_t} \exp({\rm i} \theta_2) 
- \frac{1}{y^2_{\rm d} \sqrt{1-y^4_d}} \frac{m_s}{m_b} \exp({\rm i}\theta_3) \right ] \; . \nonumber
\end{eqnarray}
\normalsize

To transform the KM matrix $V$ obtained above to a more familiar Wolfenstein 
form \cite{Wolf}, we make the following phase rotations:
\begin{eqnarray}
& & c \; \rightarrow \; c \exp({\rm i}\theta_3) \; , ~~~~~~~~
t \; \rightarrow \; t \exp[{\rm i} (\theta_3 - \varphi)] \; , \nonumber \\
&  & u \; \rightarrow \; u \exp({\rm i}\phi) \; , ~~~~~~~~
d \; \rightarrow \; d \exp({\rm i}\phi) \; , \\ 
&  &b \; \rightarrow \; b \exp(-{\rm i}\varphi) \; , \nonumber
\mlab{rotation}
\end{eqnarray}
where the phase shifts $\phi$ and $\varphi$ are defined as
\begin{eqnarray}
\phi & = & \arctan \left [ \frac{\sin\theta_2}{\cos\theta_2 ~ - ~
\displaystyle\frac{y_{\rm d} z_{\rm d}}{y_{\rm u} z_{\rm u}} \sqrt{\frac{m_c m_d}{m_u m_s}}}
\right ] \; , \nonumber \\ 
\varphi & = & \arctan \left [ \frac{\sin (\theta_3 - \theta_2)}
{\cos (\theta_3 - \theta_2) ~ - ~ \displaystyle\frac{y^2_{\rm d} 
\sqrt{1-y^4_{\rm u}}}{y^2_{\rm u} \sqrt{1-y^4_{\rm d}}} \frac{m_t m_s}{m_c m_b}}
\right ] \; . 
\mlab{phases}
\end{eqnarray}
Accordingly, we get
\begin{equation}
V \; \approx \; \left ( \begin{array}{ccc}
\displaystyle 1-\frac{1}{2} \lambda^2	& \lambda	& V_{ub} \\ \\
- \lambda	& \displaystyle 1-\frac{1}{2}\lambda^2	& A\lambda^2 \\ \\
V_{td}	& -A\lambda^2	& 1 
\end{array} \right ) \; ,
\mlab{Wform}
\end{equation}
where
\begin{eqnarray}
\lambda & = & \sqrt{y^2_{\rm u} z^2_{\rm u} \frac{m_u}{m_c} + 
y^2_{\rm d} z^2_{\rm d} \frac{m_d}{m_s} - 2 y_{\rm u} z_{\rm u} y_{\rm d} z_{\rm d} 
\sqrt{\frac{m_u m_d}
{m_c m_s}} \cos\theta_2 } \; , \nonumber \\
A\lambda^2 & = & \sqrt{\frac{y^4_{\rm u}}{1-y^4_{\rm u}} \frac{m^2_c}
{m^2_t} + \frac{y^4_{\rm d}}{1-y^4_{\rm d}} \frac{m^2_s}{m^2_b} - 
\frac{2 y^2_{\rm u} y^2_{\rm d}}{\sqrt{1-y^4_{\rm u}} \sqrt{1-y^4_{\rm d}}} \frac{m_c m_s}
{m_t m_b} \cos(\theta_3 - \theta_2)} \; ; 
\mlab{lz1}
\end{eqnarray}
and
\small
\begin{eqnarray}
V_{ub} & \approx & y_{\rm u} z_{\rm u} \sqrt{\frac{m_u}{m_c}} \left [
\frac{y^2_{\rm d}}{\sqrt{1-y^4_{\rm d}}} \frac{m_s}{m_b} 
- \frac{1}{y^2_{\rm u} \sqrt{1-y^4_{\rm u}}} \frac{m_c}{m_t} 
\exp[{\rm i}(\theta_3 - \theta_2)] \right ] \exp[{\rm i}(\theta_2 -\phi -\varphi)]
\nonumber \\ 
&  & + ~ \frac{z_{\rm d} \sqrt{1-y^4_{\rm d}}}{y_{\rm d}} \sqrt{\frac{m_d}{m_s}}
\frac{m_s}{m_b} \exp[-{\rm i} (\phi + \varphi)] \; , \nonumber \\
V_{td} & \approx & y_{\rm d} z_{\rm d} \sqrt{\frac{m_d}{m_s}} \left [
\frac{y^2_{\rm u}}{\sqrt{1-y^4_{\rm u}}} \frac{m_c}{m_t} \exp[{\rm i}(\theta_2 -\theta_3)] 
- \frac{1}{y^2_{\rm d} \sqrt{1-y^4_d}} \frac{m_s}{m_b} \right ] 
\exp[{\rm i}(\phi + \varphi)] \nonumber \\
&  & + ~ \frac{z_{\rm u} \sqrt{1-y^4_{\rm u}}}{y_{\rm u}} \sqrt{\frac{m_u}{m_c}}
\frac{m_c}{m_t} \exp[{\rm i}(\phi +\varphi -\theta_3)] \; .
\mlab{VubVtd}
\end{eqnarray}
\normalsize

\section{Discussion}

Now let us compare these results with \mref{clue}. For simplicity,
we shall drop the terms which are suppressed by $1/m_t$.
We then get
\ba
A\lambda^2 & \approx & \frac{y^2_{\rm d}}{\sqrt{1-y^4_{\rm d}}}\frac{m_s}{m_b} \; ,
\mlab{Vcb} \\
\left | \frac{V_{td}}{V_{ts}} \right |  & \approx &
\frac{z_{\rm d}}{y^3_{\rm d}}\sqrt{\frac{m_d}{m_s}} \; , 
\mlab{VtdVts} \\
\left |\frac{V_{ub}}{V_{cb}} \right |  & \approx &
\sqrt{y_{\rm u}^2z_{\rm u}^2\frac{m_u}{m_c}+\frac{z_{\rm d}^2(1-y^2_{\rm d})^2}
{y_{\rm d}^6}\frac{m_d}{m_s}
+\frac{2y_{\rm u}z_{\rm u}z_{\rm d} \left (1-y^4_{\rm d} \right )}
{y_{\rm d}^3}\sqrt{\frac{m_um_d}{m_cm_s}} \cos\theta_2} \; .
\mlab{VubVcb}
\ea
Considering the fact that \mref{clue} is a simple guess, we find that
these results together with
the expression for $V_{us}=\lambda$ given in \mref{lz1},
are quite consistent with \mref{clue}.

For the purpose of illustration, let us estimate the magnitudes of the relevant
parameters. There are  six unknown parameters 
($y_{\rm u}$, $y_{\rm d}$, $z_{\rm u}$, $z_{\rm d}$, $\theta_2$ and $\theta_3$),
and only four of them can be determined from the experimental data on $V$.
In the limit that top quark is very heavy, we see that $\theta_3$ and $y_{\rm u}/z_{\rm u}$ 
are the two parameters which remain undetermined.

(i) With the help of the experimental value $|V_{cb}| \approx 0.04$ \cite{PDG} 
and the scale-independent result $m_b/m_s \approx 34$ \cite{Narison}, 
\mref{Vcb} leads to  $y_{\rm d}\approx 0.9$.

(ii) Current data together with the unitarity of $V$ suggests $0.13 \leq |V_{td}/V_{ts}| \leq 0.35$ 
\cite{Ali}, while the magnitude of $m_s/m_d$ is allowed to be in the range 17 to 25 \cite{PDG}.  
Typically taking $m_s/m_d \approx 21$, $y_{\rm d} \approx 0.9$ and $|V_{td}/V_{ts}|\approx 0.25$,
we obtain  $z_{\rm d}\approx 0.84$  from \mref{VtdVts}. 

(iii) Taking $|V_{us}|\approx 0.22$, 
$|V_{ub}/V_{cb}|\approx 0.08$ \cite{PDG} and $m_u/m_c \approx 0.004$ \cite{Gasser}, 
we find $y_{\rm u} z_{\rm u} \approx 1.3$ and $\theta_2\approx 123^0$
from \mref{VubVcb}. Accordingly, one obtains $\phi \approx 162^0$ from \mref{phases}.

We stress that
\be
y_{\rm d} \; \approx \; 0.9 \; , ~~~~~~~~
z_{\rm d} \; \approx \; 0.84 \; , ~~~~~~~~
y_{\rm u} z_{\rm u} \; \approx \; 1.3
\ee
are in agreement with assumptions given in \mref{restriction}.
 
The Wolfenstein parameters can be written as follows. First note that 
\be
\left | \frac{V_{ub}}{V_{cb}} \right | \; \approx \; \left | \lambda \exp({\rm i} \phi) 
~ + ~ \frac{z_{\rm d}}{y^3_{\rm d}} \sqrt{\frac{m_d}{m_s}} \right | \; , 
\mlab{unit}
\ee
which is consistent with  unitarity of $V$. From \mref{unit} we deduce
\begin{equation}
\rho \; \approx \; 1 ~ + ~ \frac{z_{\rm d}}{y^3_{\rm d} \lambda}
\sqrt{\frac{m_d}{m_s}} \cos\phi \; , ~~~~~~~~
\eta \; \approx \; \frac{z_{\rm d}}{y^3_{\rm d} \lambda} 
\sqrt{\frac{m_d}{m_s}} \sin\phi \; .
\mlab{rhoeta}
\end{equation}
In the $\rho - \eta$ plane, the unitarity triangle
formed by three sides $V_{ub}^*V_{ud}$, $V^*_{cb}V_{cd}$ and $V_{tb}^*V_{td}$
can be rescaled into a simpler one with vertices $A(\rho, \eta)$, $B(1,0)$
and $C(0,0)$ \cite{PDG}. Its inner angles $\phi_1 \equiv \angle ABC$ and $\phi_2 \equiv
\angle BAC$ are approximately given as
\begin{equation}
\phi_1 \; \approx \; 180^0 - \phi \; , ~~~~~~~~
\phi_2 \; \approx \; \arctan \left [ \frac{-\sin\theta_2}{\displaystyle \cos\theta_2 +
\frac{ \left (1-y^4_{\rm d}\right ) z_{\rm d}}{y_{\rm u}z_{\rm u} 
y^3_{\rm d}} \sqrt{\frac{m_c m_d}{m_u m_s}}} \right ] \; .
\mlab{phi}
\end{equation}

Similarly, one can calculate the rephasing-invariant measure of $CP$ violation
in the $3\times 3$ quark mixing matrix (the so-called Jarlskog parameter \cite{Jarlskog}).
By use of \mref{Vresult} or \mref{Wform}, we find
\begin{equation}
J \; = \; {\rm Im}\left (V_{us}V_{cb}V^*_{ub}V^*_{cs}\right ) \; \approx \;
\frac{y_{\rm u}y_{\rm d}z_{\rm u}z_{\rm d}}{1-y^4_{\rm d}}
\sqrt{\frac{m_u}{m_c}}\sqrt{\frac{m_d}{m_s}} 
\left (\frac{m_s}{m_b}\right )^2 \sin\theta_2 \; .
\mlab{J}
\end{equation}
With the help of the rough results obtained above, we get
$J\approx 2.9\times 10^{-5}$, consistent with the present experimental expectation.

Note that the result of $V_{cb}$ obtained in \mref{Vresult} or \mref{Vcb} 
is quite different
from that predicted by the Fritzsch {\it Ansatz} (one of the simplest patterns
of the NNI mass matrices \cite{Fritzsch}):
\begin{equation}
|V_{cb}|_{\rm Fr} \; \approx \; \left | \sqrt{\frac{m_s}{m_b}} ~ - ~ 
\sqrt{\frac{m_c}{m_t}} ~ \exp({\rm i}\theta_3) \right |, 
\mlab{VcbFr}
\end{equation}
which does not agree with the experimental values of $|V_{cb}|$ 
and  the top quark mass $m_t$. 
This difference is easily understood, since \mref{VcbFr}
arises from $b=d\approx \sqrt{m_2/m_3}$ while \mref{Vcb} is guaranteed by
the condition in Eq.(12b) with $b\propto m_2/m_3$. Nevertheless, we 
find that $z=1$, assumed in the Fritzsch mass matrices, remains to be an
interesting choice. It leads to a consistent set of values for
 $V_{us}$, $V_{cd}$, $V_{ub}$ and $V_{td}$ in \mref{Vresult}.
One can conclude that in the NNI basis $d\gg b$, i.e., a large asymmetry 
between (2,3) and (3,2) elements of $M^{\prime}$, is the necessary condition
for $M^{\prime}$ to accommodate the experimental value of $V_{cb}$.
This point has also been noticed in Refs.\cite{Branco2} and \cite{Dutta},
where $d_{\rm u} \gg b_{\rm u}$, $d_{\rm d} = b_{\rm d}$ and $d_{\rm u}
= e_{\rm u} \gg b_{\rm u}$, $d_{\rm d} = e_{\rm d} \gg b_{\rm d}$ are
respectively taken. Finally it is worth mentioning that Ito and Tanimoto
\cite{Ito} have studied another parameter space of the generic NNI mass matrices
and their analytical results for $V_{cb}$, $V_{ts}$, $V_{ub}$ and $V_{td}$
are different from ours.

Without loss of generality, it is more common to construct Hermitian mass 
matrices from a theory.
If we transform $M^{\prime}$ into another weak basis in which the resultant
quark mass matrix takes the Hermitian form, then its (2,2) element
should be nonvanishing \cite{Xing},
as required by the inequality of $b$ and $d$ in $M^{\prime}$.

\section{On the scale dependence}

Note that the mass matrix $M^{\prime}$ depends upon the underlying
renormalization scale $\mu$. 
In our discussion above, we took $M^{\prime}$ defined 
at a low-energy
scale $\mu_0$ ($\leq m^{~}_Z$).
The running of $M^{\prime}$
from $\mu_0$ to an arbitrary renormalization point $\mu$ (e.g., the 
grand unification theory scale) depends on
a specific model of spontaneous symmetry breaking. In general, the
texture zeros of $M^{\prime}$, as well as the parallel structures of 
$M^{\prime}_{\rm u}$ and $M^{\prime}_{\rm d}$, may disappear after the
evolution \cite{AL}.

Of course, the KM matrix elements are scale dependent too. 
A detailed analysis has shown that $|V_{ub}|$, $|V_{cb}|$, $|V_{td}|$
and $|V_{ts}|$ have identical evolution from $\mu_0$ to $\mu$
($> \mu_0$), and the running effects of $m_u/m_c$, $m_d/m_s$, $|V_{us}|$ 
and $|V_{cd}|$ are negligibly small \cite{Babu}. 
Therefore, the approximate results for $|V_{ub}/V_{cb}|$ 
in \mref{VtdVts} and $|V_{td}/V_{ts}|$ in \mref{VubVcb} are 
independent of the renormalization scale $\mu$. 
Also, $|V_{cb}|$ and $m_s/m_b$
may have the same running behavior (dominated by the top quark Yukawa 
coupling) in the standard electroweak model
or its supersymmetric extension with small $\tan\beta$ (the ratio
of Higgs vacuum expectation values) \cite{Babu}.
In this case, the approximate formula for $|V_{cb}|$ in \mref{Vcb}
is scale independent. For large $\tan\beta$ (e.g., $\tan\beta \geq m_t/m_b$), 
however, the terms proportional to $m_c/m_t$ in \mref{Vresult} may be 
non-negligible and our approximate results in \mref{Vcb}, \mref{VtdVts}
and \mref{VubVcb} would involve larger errors. At least,
an additional evolution factor has to be introduced into
\mref{Vcb} due to the significantly different renormalization effects 
between $|V_{cb}|$ and $m_s/m_b$. One can see that the scale dependence 
of $J$ in \mref{J} is dominantly controlled by that of $m_s/m_b$.

\section{Summary}

Noticing that the KM matrix elements are numerically equal to certain
functions of quark mass ratios,
we have presented a class of {\it Ans$\it\ddot{a}$tze} for the quark mass matrices.
It is given in the NNI form where all irrelevant parameters
of the mass matrices have been removed. The main assumption is that
all hierarchy present in the KM matrix elements are due to quark mass ratios
and there is no other small parameter or delicate cancellation.

\vspace{0.4cm}
\begin{flushleft}
{\Large\bf Acknowledgements}
\end{flushleft}

One of the authors, AIS, acknowledges  Daiko Foundation for a partial support of
his research, and also ZZX is indebted to the Japan Society for the Promotion of Science
for its financial support.

\end{document}